\documentclass[draft,tightenlines,nofootinbib,preprint,aps,eqsecnum]{revtex4}

\newcommand{\beq}{\begin{equation}}
\newcommand{\eeq}{\end{equation}}
\newcommand{\bea}{\begin{eqnarray}}
\newcommand{\eea}{\end{eqnarray}}

\begin{document}

\title{On Relativistic Entanglement and Localization of Particles and on
their Comparison with the Non-Relativistic Theory}
\author{Horace W. Crater}
\affiliation{The University of Tennessee Space Institute\\
Tullahoma, TN 37388 USA\\
E-mail: hcrater@utsi.edu}
\author{Luca Lusanna}
\affiliation{Sezione INFN di Firenze\\
Polo Scientifico\\
Via Sansone 1\\
50019 Sesto Fiorentino (FI), Italy\\
E-mail: lusanna@fi.infn.it}

\begin{abstract}
We make a critical comparison of relativistic and non-relativistic classical
and quantum mechanics of particles in inertial frames as well of the open
problems in particle localization at both levels. The solution of the
problems of the relativistic center of mass, of the clock synchronization
convention needed to define relativistic 3-spaces and of the elimination of
the relative times in the relativistic bound states leads to a description
with a decoupled non-local (non-measurable) relativistic center of mass and
with only relative variables for the particles (single particle subsystems
do not exist). We analyze the implications for entanglement of this
relativistic spatial non-separability not existing in non-relativistic
entanglement. Then we try to reconcile the two visions showing that also at
the non-relativistic level in real experiments only relative variables are
measured with their directions determined by the effective mean classical
trajectories of particle beams present in the experiment. The existing results
about the non-relativistic and relativistic localization of particles and atoms
support the view that detectors only identify effective particles following this
type of trajectories: these objects are the phenomenological emergent aspect
of the notion of particle defined by means of the Fock spaces of quantum field theory.

\end{abstract}

\today

\maketitle

\vfill\eject

\section{Introduction}

The classical mechanics (CM) of Newton gives a deterministic
description of objects (particles, bodies) supposed to have a reality in an
inertial frame of the Galilei space-time centered on an inertial
mathematical observer playing no dynamical role beyond defining Cartesian
coordinates. This space-time is assumed to be a given background container
of the real objects, whose world-lines are described in terms of an absolute
notion of time. At each instant there is an absolute Euclidean 3-space where
the objects are localized. The inertial frames are connected by the
transformations of the  Galilei group. This description can be
extended to non-inertial frames centered on mathematical accelerated
observers.

\bigskip

This realistic description of the world-lines of particles is preserved in
special relativity (SR). \ However, now they are described in the inertial
frames of Minkowski space-time centered on inertial mathematical
relativistic observers and the  Poincar\'{e} group describes the
transformations connecting the inertial frames. However, in SR there is no
notion of absolute time and of absolute 3-space: only the whole Minkowski
space-time is absolute and only the conformal structure (i.e. the light-cone
describing the locus of incoming and outgoing radiation in every point) has
an intrinsic meaning. As a consequence, we must introduce a \textit{%
convention of clock synchronization} to define an instantaneous 3-space,
whose definition is needed to formulate the Cauchy problem for wave
equations like the Maxwell's ones.

\bigskip

In this Introduction we give an outline of  the main problems to be faced in the
definition of relativistic classical mechanics (RCM) and relativistic quantum mechanics
(RQM). This will clarify the context which gave origin to this paper and to its
results and implications.

\bigskip

Usually RCM is formulated in inertial
frames, whose Euclidean 3-spaces are defined by Einstein's convention
\footnote{The inertial observer A sends a ray of light at $x_{i}^{o}$ towards the (in
general accelerated) observer B; the ray is reflected towards A at a point P
of B world-line and then reabsorbed by A at $x_{f}^{o}$; by convention P is
synchronous with the mid-point between emission and absorption on A's
world-line, i.e. $x_{P}^{o}=x_{i}^{o}+{\frac{1}{2}}\,(x_{f}^{o}-x_{i}^{o})={%
\frac{1}{2}}\,(x_{i}^{o}+x_{f}^{o})$. This convention selects the Euclidean
instantaneous 3-spaces $x^{o}=ct=const.$ of the inertial frames centered on
A. However, if the observer A is accelerated, the convention can breaks down
due the possible appearance of coordinate singularities.}. Only with this
convention does the 1-way velocity of light between two observers (it
depends on how their clocks are synchronized) coincide with the 2-way
velocity of light $c$ of an inertial observer (it replaces the unit of
length in relativistic metrology \cite{1}).

\bigskip

However this description of RCM is still incomplete for interacting systems
due to the following problems:

1) There is not a unique notion of relativistic center of mass of a system
of particles like in Newtonian mechanics;

2) There is the problem of the elimination of relative times in relativistic
bound states (time-like excitations are not seen in spectroscopy);

3) It is highly non trivial to find the explicit form of the Poincar\'e
generators (especially the Lorentz boosts) for interacting particles in the
instant form of dynamics;

4) There is no accepted global formulation of non-inertial frames without
the pathologies of the rotating disk and of Fermi coordinates.

\bigskip

Recently a solution to all these problems has been given in Refs.\cite{2,3,4}
(see also the review in Ref.\cite{5}). As sketched in Section II, the main
differences from non-relativistic CM are the \textit{non-local nature} of
the relativistic collective variables proposed for the relativistic center
of mass (implying their non-measurability with local measurements) and a
\textit{spatial non-separability} of the particles, which must be described
by means of suitable Wigner-covariant relative 3-variables.

\bigskip

This formulation of RCM allows one to get a consistent definition of
RQM of particles with an associated notion
of relativistic entanglement as an extension of non-relativistic quantum
mechanics (NRQM) avoiding all the known relativistic pathologies. This was
done in Ref.\cite{6}. This framework for RCM has been extended to classical field theory (CFT) in  Refs.\cite
{3,5} both for classical fields and fluids, but the  extension of the approach to quantum field theory
(QFT) has still to be done.

\bigskip

Unlikely from the transition from Galilei space-time to the Minkowski
one, the transition from  CM to NRQM can be done only in an
operational way due to the big unsolved foundational problems of NRQM (see
for instance Ref.\cite{7}) \footnote{Let us remark that for many physicists the absence of experimental facts in contrast with QM is an indication that the foundational problems are fake problems of philosophical type. See
for instance Ref.\cite{8}. Our attitude is neither operational nor foundational: we try to understand some aspects of the transition from the quantum to the classical regime following Bohr's viewpoint.}. The main problem is that the notion of {\it reality} of a classical particle and of its properties cannot be extended to NRQM as
shown

A) by the EPR experiment (see Ref.\cite{9} for a review) and the violation
of Bell's inequalities (no local realistic hidden variable explanation; see
Ref.\cite{10} for the status of experiments);

B) by the Kochen-Specker theorem \cite{11} (no non-contextual explanation of
the properties of a quantum system);

C) by the probabilistic Born's rule for the unique outcomes of measurements
(in a random way a unique value is obtained for the observable, describing a
property of the state of the quantum object under investigation; but a
repeated measurement on an identical state can give other values, i.e. we
cannot speak of a property of the quantum system but only of quantity which
takes value depending on the context randomly).

\bigskip

As a consequence it is not clear which is the meaning of the localization of
a quantum particle even having taken into account the Heisenberg uncertainty
relations (see Ref.\cite{12} for the problem of simultaneous measurements of
position and momentum).

\medskip

On one side we have the mathematical theory of NRQM with the unitary
evolution of the wave function, but there is no consensus on whether this
wave function describes the given quantum system or is only an information
on a statistical ensemble of such systems. Moreover it could be that only
the real-valued density matrix, i.e. the statistical operator determining the probabilities,
makes sense and not the complex-valued wave function (only a mathematical tool). There is no accepted interpretation
for the theory of measurements going beyond the non-unitary collapse of the
wave function in an instantaneous idealized Von Neumann measurement of a
self-adjoint operator describing some mathematical observable property of
the quantum system.

\medskip

In experiments we have macroscopic semi-classical objects as source and
detectors of quantities named quantum particles (or atoms) and the results
shown by the pointers of the detectors are the end point of a macroscopic
(many-body) amplification of the interaction of the quantum object with some
microscopic constituent of the detector (for instance an $\alpha $ particle
interacting with a water molecule followed by the formation of a droplet as
the amplification allowing detection of the particle trajectory in bubble
chambers). Usually one invokes the theory of decoherence \cite{13,14} with
its uncontrollable coupled environment for the emergence of robust classical
aspects explaining the well defined position of the pointer in a measurement.

\bigskip

Also recently Haag \cite{15} said that the deterministically propagating
(due to the Schroedinger equation) pure state of a quantum particle has no
objective significance and do not represent a real phenomenon. The only
relevant notion are the \textit{events}, namely a set of mutually exclusive
possibilities with an associate probability assignment for the outcomes of
operationally defined measurements. To identify events in particle physics
we need not only the beams of incoming particles but also an effective microscopic
description of the interaction of the particle with some constituent of the
macroscopic detector. This means that we need a localization region in
space-time and a description of the detector (what it is supposed to measure
with a microscopic interaction then suitably amplified). Then we need a
principle of random realization to justify the unique outcome of each
measurement, whose repetition gives results distributed according to the
Born rule. In this ensemble interpretation only the events, containing not
only the quantum particle but also the measuring apparatus, have some kind
of reality: due to decoherence things happens "as if" a semi-classical
particle interacts with  semi-classical constituents of the detector.

\bigskip

A first step to face all these problems (always in an ensemble
interpretation) is presented in Ref.\cite{16}. It is an approach more
general than decoherence but limited to spin systems (the only ones where
many-body calculations can be explicitly made). One considers the
interacting object as an \textit{open quantum subsystem} \cite{17} of a
macroscopic many-body system (system plus detector plus environment) with
the induced non-unitary stochastic behavior (even without going to the
thermodynamic limit): now time scales for the various phases of the
measurement and aspects of decoherence can be explicitly evaluated.

\medskip

The described state of affairs is in accord with Bohr's point of view
according to which we need a classical description of the experimental
apparatus. It seems that all the realizable experiments must admit a
quasi-classical description not only of the apparatus but also of the
quantum particles: they are present in the experimental area as classical
effective  particles with a mean trajectory and a mean value of 4-momentum
(measured with time-of-flight methods).

\medskip

As shown in Ref.\cite{18}, if one assumes that the wave function describes
the given quantum system (no ensemble interpretation), the statement of Bohr
can be justified by noting that the wave functions used in the preparation
of particle beams (semi-classical objects with a mean classical trajectory
and a classical mean momentum determined with time-of-flight methods) are a
special subset of the wave functions solutions of the Schroedinger equation
for the given particles. Their associated density matrix, pervading the
whole 3-space, admits a \textit{multi-polar expansion around a classical
trajectory} having \textit{zero dipole}. This implies that in this case the
equations of the Ehrenfest theorem give rise to the Newton equations for the
Newton trajectory (the monopole; it is not a Bohmian trajectory) with a
classical force augmented by forces of a quantum nature coming from the
quadrupole and the higher multipoles (they are proportional to powers of the
Planck constant). As shown in Ref.\cite{18} the mean trajectories of the
prepared beams of particles and of the particles revealed by the detectors
are just these classically emerging Newton trajectories impled by the
Ehrenfest theorem for wave functions with zero dipole. Also all the
intuitive descriptions of experiments in atomic physics are compatible with
this emergence of classicality. In these descriptions an atom is represented
as a classical particle delocalized in a small sphere, whose origin can be
traced to the effect of the higher-multipole forces in the emerging Newton
equations for the atom trajectory. The wave functions without zero dipole do
not seem to be implementable in feasible experiments. No explanation is
given of the probabilistic Born rule, but it is suggested that the random
unique outcomes have a quasi-classical localization given by these Newton
trajectories. \bigskip

In this paper we want to focus on the problems connected with the
localization of particles both at the relativistic and non-relativistic
levels and both at the classical and quantum levels QFT included. We identify the existing proposals for position measurements and we analyze the existing theoretical problems in this area (usually they are not well known to many researchers).  The comparison with the experimental status of
particle and atom localization will help to clarify whether it is possible
to measure the non-relativistic center of mass of a system or whether it is
non-measurable like the relativistic collective variables. Then the results
on localization will be used to clarify the connection between relativistic
and non-relativistic entanglement and what can be seen in the experiments.

We hope that collecting and gluing together results on these topics coming
from usually non mutually interacting communities will be helpful for
researchers approaching the quickly developing areas of mesoscopic physics,
atomic and molecular physics, atomic clocks and space physics, quantum
information, teleportation,....

\medskip

In Section III we make a review of the problems in the localization of
particles both at the relativistic (Subsection A) and at the non-relativistic
(Subsection B) level; then in Subsection C we look at the notion of particle in QFT and
to its problems. In Section IV we show the differences between non-relativistic and relativistic
entanglement in a two-body case induced by the relativistic spatial
non-separability forbidding the identification of subsystems. In Section V
we study the preparation and detection of particles in experiments and we
propose a reconciliation of the non-relativistic and relativistic visions
valid for all practical purposes. In the final Section there are concluding
remarks and a list of open problems.

\section{Review of Relativistic Classical and Quantum Mechanics}

The new formulation of RCM \cite{2,3,5} and  of  a consistent RQM \cite{6} makes
use of the 3+1 point of view to build a theory of global non-inertial frames centered on arbitrary time-like
observers \cite{2,5}. This is done by giving the world-line of the time-like
observer and a nice foliation of Minkowski space-time with non-intersecting
space-like Riemannian 3-spaces, all tending to the same space-like
hyper-plane at spatial infinity. Moreover, one uses the radar 4-coordinates $%
(\tau ;\sigma ^{r})$, i.e. an arbitrary monotonically increasing function of
the proper time of the atomic clock carried by the observer and curvilinear
3-coordinates $\sigma ^{r}$ centered on the observer for the 3-spaces.
CFT may be reformulated in this framework by using fields knowing the clock synchronization convention.

\medskip

Both the knowledge of the whole world-line of an arbitrary time-like
observer and of nice foliation with 3-spaces of Minkowski space-time are
\textit{non-factual}  notions. The observer is a purely
mathematical entity carrying a clock, an idealization of a physical atomic
clock carried by a dynamical observer. The foliation is the mathematical
idealization of a physical protocol of clock synchronization. Actually the
physical protocols (think of GPS) can establish a clock synchronization
convention only inside future light-cone of the physical observer defining
the local 3-spaces only inside it. However to be able to formulate the
Cauchy problem for field equations and to have predictability of the future,
due to the theorem on the existence and unicity of solution of partial
differential equations we have to extend the convention outside the
light-cone \footnote{%
As far as we know the theorem on the existence and unicity of solutions has
not yet been extended starting from data given only on the past light-cone.}%
.. Once we have given the Cauchy data on the initial Cauchy surface (an
unphysical process), we can predict the future with every observer receiving
the information only from his/her past light-cone (retarded information from
inside it; electromagnetic signals on it). \medskip

For non-relativistic observers the situation is simpler, but the non-factual
need of giving the Cauchy data on a whole initial absolute Euclidean 3-space
is present also in this case for non-relativistic field equations like the
Euler equation for fluids.

\subsection{Relativistic Classical Mechanics}

As shown in Ref.\cite{2} in this framework the description of isolated
systems can be done with an action principle (the \textit{parametrized
Minkowski theories} for particles, fields, strings, fluids) implying that
the transition among non-inertial frames is described by gauge
transformations (so that only the appearances of phenomena change, not the
physics) and allowing one to define the energy-momentum tensor and then the
Poincar\'{e} generators of the system. \medskip

Inertial frames are a special case of this theory having Euclidean 3-spaces.
For isolated systems there is a special family of inertial systems, the
\textit{intrinsic rest frames}, in which the space-like 3-spaces are
orthonormal to the conserved time-like 4-momentum of the isolated system. At
the Hamiltonian level it turns out that every isolated system can be
described by a decoupled canonical non-covariant relativistic center of mass
(whose spatial part is the classical counterpart of the Newton-Wigner
position operator). \ Such a system carries a pole-dipole structure, namely
an internal 3-space with a well defined total invariant mass $M$ and a total
rest spin $\vec{S}$ and a well defined realization of the Poincar\'{e}
algebra (\textit{\ the external Poincar\'{e} group} for a free point
particle, which we identify with the \textit{\ external center of mass},
whose mass and spin are its Casimir invariants describing the matter of the isolated
system in a global way). The internal rest 3-space, named Wigner 3-space, is
defined in such a way that it is the same for all the inertial rest frames
and its 3-vectors are Wigner spin-1 3-vectors \cite{19}, so that the
covariance under Poincar\'{e} transformations is under control. The
particles of the isolated system, all having the same time of the given
3-space, are identified by this type of Wigner-covariant 3-vectors (see
Refs.[3,5] for the description of fields).

\medskip

As shown in Refs.\cite{6,20} the canonical non-covariant (a pseudo 4-vector)
relativistic center of mass, the non-canonical covariant (a 4-vector)
Fokker-Price center of inertia and the non-canonical non-covariant (a pseudo
4-vector) M$\o $ller center of energy are the \textit{only three
relativistic collective variables} which can be built only in terms of the
Poincar\'{e} generators of an isolated system  so that they depend only on
the system and nothing external to it. All of them collapse onto the Newton
center of mass of the system in the non-relativistic limit \footnote{%
It is of interest that the three properties of the non-relativistic center
of mass, namely i) a position associated with the spatial mass distribution
of the constituents ii) it transformation under rotations as a three \
vector and iii) together with the total momentum being canonical variables,
have their respective relativistic counterparts taken up by the Moller
non-covariant, non-canonical center of energy $R^{\mu }(\tau )$ , the
covariant but non-canonical Fokker Pryce center of inertia \ $Y^{\mu }(\tau )
$ and the canonical, but non-covariant center of mass $\tilde{x}^{\mu }(\tau
).$ \ }.

\medskip

As shown in Refs.\cite{2,4} in the Wigner 3-space there is another
realization of the Poincar\'e algebra (\textit{the internal Poincar\'e group}%
) built with the rest 3-coordinates and 3-momenta of the matter of the
isolated system starting from its energy-momentum tensor: the internal
energy is the invariant mass $M$ (the Hamiltonian inside the rest 3-space)
and the internal angular momentum is the rest spin $\vec S$. Since we are in
rest frames the internal 3-momentum must vanish. Moreover, to avoid a double
counting of the center of mass, the internal center of mass, conjugate to
the vanishing 3-momentum, has to be eliminated: this is done by fixing the
value of the internal Poincar\'e boost. If we put it equal to zero, this
implies that the time-like observer has to be an inertial observer
coinciding with the non-canonical 4-vector describing the Fokker-Price
center of inertia of the isolated system. Therefore the internal realization
of the Poincar\'e algebra is unfaithful and inside the Wigner rest 3-spaces
the matter is described by \textit{relative} 3-positions and 3-momenta.
\medskip

The world-lines of the particles (and their 4-momenta) are derived notions,
which can be rebuilt given the relative 3-coordinates, the time-like
observer (for instance the Fokker-Price center of inertia) and the axes of
the inertial rest frame \cite{4}. They are described by 4-vectors $%
x^{\mu}(\tau)$, which however are not canonical like in most of the
approaches: there is a classical \textit{non-commutative structure} induced by the
Lorentz signature of Minkowski space-time \cite{4,6}.

\medskip

As shown in Ref. \cite{3} these three variables can be expressed as known
functions of the Lorentz scalar rest time $\tau $, of canonically conjugate
Jacobi data (frozen (fixed $\tau =0)$ Cauchy data) $\vec{z}=Mc\,{\vec{x}}
_{NW}(0)$, $\vec{h}=\vec{P}/Mc$, (${\vec{x}}_{NW}(\tau )={\vec{\tilde{x}}}
(\tau )$ is the standard Newton-Wigner 3-position; $P^{\mu }$ is the
external 4-momentum), and of the invariant mass $M$ and rest spin $\vec{S}$.
The external Poincar\'{e} generators are then expressed in terms of these
variables. \medskip

As said in Ref.\cite{6}, since the three relativistic collective variables
depend on the internal Poincar\'e generators $M$ and $\vec S$, which are
conserved integrals  of suitable components of the energy-momentum
tensor of the isolated system over the whole rest 3-space, they are \textit{%
non-local} quantities which cannot be determined with local measurements.

\subsection{Relativistic Quantum Mechanics}

The use of $\vec{z}$ avoids taking into account the mass spectrum of the
isolated system at the quantum kinematical level and allows one to avoid the
Hegerfeldt theorem (the instantaneous spreading of wave packets with
violation of relativistic causality) in RQM \cite{6}.

\bigskip

Besides these non-local features in RQM there is an intrinsic \textit{%
spatial non-separability} forbidding the identification of subsystems at the
physical level and generating a notion of relativistic entanglement very
different from the non-relativistic one. \medskip

In order to exhibit these two properties, let us consider a quantum two-body
system. In non-relativistic quantum mechanics (NRQM) its Hilbert space can
be described in the three following unitarily equivalent ways \cite{6}: A)
as the tensor product $H=H_{1}\otimes H_{2}$, where $H_{i}$ are the Hilbert
spaces of the two particles (separability of the two subsystems as the
zeroth postulate of NRQM); B) as the tensor product $H=H_{com}\otimes H_{rel}
$, where $H_{com}$ is the Hilbert space of the decoupled free Newton center
of mass and $H_{rel}$ the Hilbert space of the relative motion (in the
interacting case only this presentation implies the separation of variables
in the Schroedinger equation); C) as the tensor product $H=H_{HJcom}\otimes
H_{rel}$, where $H_{HJcom}$ is the Hilbert space of the frozen Jacobi data
of the Newton center of mass (use is made of the Hamilton-Jacobi
transformation). \medskip

Each of these three presentations gives rise to a different notion of
entanglement due to the different notion of separable subsystems. As shown
in Ref. \cite{19} other presentations are possible in NRQM: in each
presentation there is a different notion of separable or entangled pure
state (the same is true in the mixed case).

\medskip

As shown in Ref.\cite{6}, at the relativistic level the elimination of the
relative times of the particles (they are defined in a 3-space with a
definite value of time) and the treatment of the relativistic collective
variables allows \textit{only the presentation C)}, i.e. $H=H_{HJcom}\otimes
H_{rel}$ with $H_{HJcom}$ being the Hilbert space associated to the
quantization of the canonically conjugate frozen Jacobi data $\vec{z}$ and $%
\vec{h}$ and $H_{rel}$ is the Hilbert space of the Wigner-covariant relative
3-coordinates and 3-momenta. Therefore only the frozen relativistic 3-center
of mass and the set of all the relative variables are the admissible
separable relativistic subsystems in RQM. Already at the classical level the
subsystems particle 1 and particle 2 (without relative times) are only
defined in the \textit{un-physical} rest 3-space, which is how one terms the
3-space before adding the rest-frame conditions eliminating the internal
3-center of mass and its 3-momentum. In contrast, the physical space is the
one with these constraints imposed. The rest-frame conditions, defining the
physical variables, destroy the separability of the particles leaving only
relative variables. In this framework there are no problems with the
treatment of relativistic bound states \footnote{%
If one considers the tensor product $H_{1}\otimes H_{2}$ of two massive
Klein-Gordon particles most of the states will have one particle allowed to
be the absolute future of the other due to the lack of restrictions on the
relative times. Only in S-matrix theory is this irrelevant since one takes
the limit for infinite future and past times.}. \medskip

Let us remark that instead of starting from the physical Hilbert space
containing the frozen Jacobi data, one could first define an un-physical
Hilbert space containing the Jacobi data and the 3-position and 3-momenta of
the particles (in it we have the same kind of separability as in the
presentation A) of NRQM) and then define the physical Hilbert space by
imposing the rest-frame conditions at the quantum level with the
Gupta-Bleuler method. However there is the risk to get an inequivalent
quantum theory due to the complex form of the internal boosts.

\subsection{Classical and Quantum Field Theory}

Given a 3+1 splitting associated with a time-like observer using radar 4-coordinates $\sigma^A = (\tau, \sigma^r)$
we can rebuild the Cartesian coordinates of an inertial observer of Minkowski space-time with a
coordinate transformation $\sigma^A\, \rightarrow\, x^{\mu} = z^{\mu}(\tau, \sigma^r)$ with $z^{\mu}(\tau, 0) =
x^{\mu}(\tau)$ being the world-line of the time-like observer. The functions $z^{\mu}(\tau, \sigma^r)$ describe
the embedding of the instantaneous 3-spaces $\Sigma_{\tau}$ in Minkowski space-time.

\medskip

Given a classical field, for instance the Klein-Gordon field $\tilde \phi(x^{\mu})$, its reformulation as a field knowing the clock synchronization convention is $\phi (\tau, \sigma^r) = \tilde \phi(x^{\mu} = z^{\mu}(\tau, \sigma^r))$. These are the fields used in parametrized Minkowski theories \cite{2}.

\medskip

As shown in Ref.\cite{3} for the case of particles plus the electro-magnetic field, at the classical level one can
define the relativistic external center of mass and the relative variables for these fields and find the rest-frame
conditions eliminating the internal center of mass. In atomic physics this allows to avoid pathologies like
the Haag theorem (non existence of the interaction picture in QFT) and to follow the evolution of atoms in the
interaction region for finite times taking into account the relativistic properties of non-separability and non-locality. The extension of these results to QFT is highly non-trivial, because at the classical level one uses
variables of the action-angle type for which no consistent quantization exists. The alternative is to quantize
the standard variables and to try to impose the quantum rest-frame conditions with the Gupta-Bleuler methods. In any case a consistent quantization along these lines would lead to a non-local QFT due to the relativistic preperties of non-separability and non-locality.

\section{Localization of Particles}

Both NRQM and RQM are defined on a fixed space-time structure, the Galilei
and Minkowski space-times respectively. More exactly they are defined in the
\textit{inertial} frames of these space-times, because the extension to
non-inertial frames is still an open problem \footnote{
At the classical level, the framework described in Section II for the
definition of parametrized Minkowski theories leads to the theory of global
non-inertial frames (see Ref.\cite{2}). However till now only the
quantization of particles in relativistic rotating frames and its
non-relativistic limit have been studied (see Refs.\cite{22}).}. \medskip

This spatio-temporal point of view is presupposed to the postulates of
quantum mechanics (QM) and to each possible interpretation of it. It is only
at the level of Einstein general relativity, where the metric structure of
space-time and space-time itself become dynamical, that this scheme breaks
down opening the basic problem of getting a consistent theory of quantum
gravity conciliating QM and gravitation (such a problem does not exist for
Newtonian gravity, which is defined in Galilei space-time).

\medskip

If we start with this space-time oriented point of view, QM is defined in
the Euclidean 3-spaces of the inertial frames of either Galilei or Minkowski
space-time. This implies that \textit{the coordinate representation has a
privileged kinematical and descriptive status} among all possible bases in
the Hilbert space of quantum systems. Since all the experiments are
localized in space-time, it is important to consider always the trajectories
of the carriers of quantum properties (like spin or qubits or other quantum
numbers) and not to treat the quantum systems independently from their
localization in the space-time. See the second part of the review paper \cite%
{23} for the relevance of Lorentz transformations for creating entanglement
between spin and momentum degrees of freedom.

This privileged status of the coordinate representation coming from the
spatio-temporal interpretation is different (but is reinforced) by the
existence of a natural selection of robust positional bases of pointer
states for the apparatuses appearing in the \textit{de-coherence} approach
to QM with its dominant role of the environment in the description of
entanglement (see Refs. \cite{13,14}).

\bigskip

Since the relativistic spatial non-separability forbidding the
identification of the subsystems of the given quantum system is a
consequence of defining relativistic collective position variables, one has
to face the open problem of \textit{position measurements} in QM. While most
of the mathematical properties of quantum systems are based on instantaneous
precise measurements of self-adjoint bounded operators (the observables)
with a discrete spectrum, whose treatment requires projection operators (or
projection valued measures, PVM), position operators are usually described
by self-adjoint unbounded operators without normalized position eigenvalues
(usually one uses the improper Dirac kets $|\vec{x}\,>$, sharp eigenstates
of the position operator, satisfying $<\vec{x}|\vec{y}>=\delta ^{3}(\vec{x}-%
\vec{y})$). However, as noted in Ref.\cite{13}, to be able to describe the
standard (even if questionable) postulate of the non-unitary collapse of the
wave function one needs position wave functions with a finite support
(inside the apparatus) to avoid the necessity of using arbitrary strong
couplings and arbitrarily large amounts of energy to make an arbitrarily
precise measurement of position. As a consequence, the notion of \textit{%
unsharp} positions with bad localization has emerged (see Refs.\cite{24} for
the theory of projection operator valued measures, POVM). The results of
measurements of a POVM give imprecise information of stochastic type on the
localization of particles (see for instance Ref.\cite{25} for continuous
quantum position weak measurements).

\medskip

We will now describe some of the existing problems with the notions of
position and localization both at the relativistic and non-relativistic
levels, having in mind the following question: "Is the center-of-mass
position measurable"?

\subsection{The Relativistic Case}

In the relativistic case there are two types of problems, one at the
classical level, the other at the quantum level. \bigskip

$\alpha $) \textit{M$\o $ller non-covariance world-tube} \cite{26}. As we
have said, in each relativistic inertial frame one has the world-line of the
Fokker-Price non-canonical, covariant center of inertia $Y^{\mu }(\tau )$ of
the isolated system and different pseudo-world-lines for the non-covariant,
canonical 4-center of mass ${\tilde{x}}^{\mu }(\tau )$ and for the
non-covariant, non-canonical M$\o $ller center of energy $R^{\mu }(\tau )$.
If in a given inertial frame we consider the positions of ${\tilde{x}}^{\mu
}(\tau )$ and $R^{\mu }(\tau )$ corresponding to every possible inertial
frame, we get a tube centered on $Y^{\mu }(\tau )$ ( with ${\tilde{x}}^{\mu
}(\tau )$ always lying between $Y^{\mu }(\tau )$ and $R^{\mu }(\tau )$). The
invariant radius of the tube is determined by the two Casimirs invariant
mass $M$ and rest spin $\vec{S}$: $\rho =|\vec{S}|/Mc$. As said in Ref.\cite%
{3}, this classical intrinsic radius is a \textit{non-local effect of the
Lorentz signature } of Minkowski space-time absent in Euclidean spaces and
delimits the place of the non-covariant effects (the pseudo-world-lines)
connected with the relativistic collective variables. These effects are not
classically detectable because the M$\o $ller radius is of the order of the
Compton wavelength of the isolated system: an attempt to test its interior
would mean to enter in the quantum regime of pair production. The M$\o $ller
radius $\rho $ is also a remnant of the energy-conditions of general
relativity in the flat Minkowski space-time: if a body has its material
radius less than its M$\o $ller one, then there is some inertial frame in
which the energy density of the body is not positive definite even if the
total energy is positive \cite{26}. \medskip

Therefore the Compton wavelength is the best theoretical approximation for
the localization of a classical massive particle.

\bigskip

$\beta $) {Newton-Wigner operator}. As found in Ref.\cite{6}, at the quantum
level the spatial component ${\vec{\tilde{x}}}$ of the canonical
non-covariant center of mass ${\tilde{x}}^{\mu }=({\tilde{x}}^{o};{\vec{%
\tilde{x}}})$ becomes the Newton-Wigner position operator \cite{27}, whose
eigenfunctions are wave functions with infinite tail and a mean width around
the eigenvalue of the order of the Compton wavelength. Therefore also at the
quantum level there is bad localization.

\medskip

In Refs.\cite{28} it is said that we cannot consider the Newton-Wigner
operator a self-adjoint operator (in the framework of quantum field theory
it is neither a local nor a quasi-local operator) but at best a \textit{%
symmetric} operator \footnote{
An operator $A$ in an infinite dimensional Hilbert spaces is said to be
symmetric if $\langle Ay|x\rangle =\langle y|Ax\rangle .$ Such operators are
not diagonalizable and therefore describe real degrees of freedom which
display a form of "unsharpness" or "fuzzyness".}. See Refs.\cite{29} for an
approach to \textit{fuzzy localization} based on the use of certain types of
symmetric operators.

\medskip

Let us remark that, whichever point of view is chosen for the position
operator, the generators of the Poincar\'{e} algebra (in particular the
Lorentz ones) of isolated systems must be described by self-adjoint
operators as it is usually assumed. This implies that the 4-momentum
operators must be self-adjoint operators.

\medskip

In the approach presented in Section II these problems appear only for the
external non-covariant canonical center of mass described by the Jacobi data
$\vec{z}$. While its conjugate variable $\vec{h}$ must be taken as a
self-adjoint operator, it is a totally open problem how to quantize $\vec{z}$
and whether one has to introduce super-selection rules \cite{30} either
forbidding its measurability or at least forbidding the possibility of
making center-of-mass wave packets (only plane waves with fixed eigenvalue
of $\vec{h}$ allowed).

\medskip

However, particle physics experiments utilize beams of particles with a mean
4-momentum and localized around a classical trajectory pointing to the
experimental area. Therefore the description of particle beams requires well
picked wave packets in momentum space with also a good localization in the
3-space.

\subsection{Non-Relativistic Case}

In standard NRQM the position of particles are usually described by
self-adjoint unbounded operators and usually one says that there are only
experimental problems with localization of particles and atoms.

\medskip

However recently in the framework of the theory of measurements based on
POVM there was a revisit of the problem by extending the Wigner-Araki-Yanase
theorem \cite{31} from bounded (like angular momentum) to unbounded
(position) operators. The theorem says that given a conserved quantity
(additive over the system plus apparatus), then a discrete self-adjoint
operator non commuting with the conserved quantity does not admit perfectly
accurate and repeatable measurements. In Refs.\cite{32} it was shown that
generically in an isolated two-body system with conserved momentum the
conjugate center-of-mass operator (and also the absolute positions of the
two particles) are \textit{unsharp}. Unsharp positions are different from
the un-determination of symmetric operators, but the final result is the
same. On the other hand, there is no problem with the relative position
variable. \medskip

\medskip

At the experimental level the previous statements have been confirmed also
in Refs.\cite{33}, where it is shown that it is only possible to measure
mutual relative positions of atoms. Regarding their absolute positions the
best localization of atoms which can be realized is at the level of hundred
of nanometers \cite{34,35}, much higher than the atom Compton wavelength.

\bigskip

In conclusion at every level we have indications that the absolute position
of massive particles can be determined only with a precision most probably
much greater that the Compton wavelength of the particle, as it happens with
the radius of the macroscopic tracks of particles in bubble chambers.
Therefore also in the non-relativistic case it seems that there are problems
with the localization of the center of mass of isolated quantum systems: one
can only say that effective atoms are inside the size of the apparatus.

As a consequence this state of affairs together with the results of Ref.\cite{33} point
to the same picture as in RQM: a non-measurable center of mass plus non-separable relative motions.
\bigskip

Let us also remark that the standard treatment of non-relativistic particles,
with its notion of separability of subsystems, ignores the fact that to take
into account electro-magnetic interactions one has to use a $1 / c$
approximation of QED below the threshold of pair production. Only in the
limit $c \rightarrow\, \infty$ one has an irreversible contraction of the
Poincar\'e algebra to the Galilei one. Therefore atomic physics needs a
relativistic formulation like the one of Ref.[5] even when the particles
have non-relativistic velocities: as just said this picture is emerging
already in NRQM. However the quantization of the
electromagnetic field in the rest frame, with the rest-frame conditions
implemented, has still to be done.

\bigskip

\medskip

Finally notions like the Planck length, dimensionally relevant when one takes into account gravity,  are completely outside the existing experimental level.

\subsection{The Notion of Particle in Quantum Field Theory}

Often it is said that RQM of particles is an irrelevant theory, because relativistic particles have to be described by QFT, which solves the problem of negative energies with anti-particles and allows pair production. This is an ambiguous statement. Firstly the description of relativistic bound states requires a transition from exact QFT equations (like the Bethe-Salpeter one) to effective RQM ones, valid below the threshold of pair production. Secondly the existing notion of particle in QFT can be done only for free fields and is subject to criticism as it can be seen from the (philosophically oriented, but mathematically relevant) papers of Ref.\cite{36} \footnote{In the interacting case one looses the control on the mass shell condition of the interacting particles. Let us remark that in formulation of RCM of Refs.\cite{2,3,4} the mass-shell condition is a derived property and depends upon the interactions.}. Finally the standard definition of particles by means of the Foc
 k space gives rise to completely delocalized objects (plane waves to be used in the in- and out-states of scattering theory). Moreover, the definition of positive- and negative-energy particles requires the existence of a time-like Killing vector of the space-time and of a suitable vacuum state (so that in curved space-times without Killing vectors one has to use algebraic QFT, where no sound definition of particle exists \cite{37}).

\medskip

Here we will consider only the massive Klein-Gordon uncharged quantum field in an inertial frame of Minkowski space-time re-expressed in the radar 4-coordinates of an inertial 3+1 splitting with Euclidean 3-spaces. It has the expression

\bea
 \hat \phi(\tau, \vec \sigma) &=& \int {{d^3k}\over {(2 \pi)^3\, 2 \omega_k}}\, \Big[e^{- i\, (\omega_k\, \tau - \vec k \cdot \vec \sigma)}\, \hat a(\vec k) + e^{i\, (\omega_k\, \tau - \vec k \cdot \vec \sigma)}\, {\hat a}^{\dagger}(\vec k)\Big],\nonumber \\
 {}&&\nonumber \\
 &&[\hat a({\vec k}_1), \hat a({\vec k}_2)] = [{\hat a}^{\dagger}({\vec k}_1), {\hat a}^{\dagger}({\vec k}_2)] = 0,\qquad [\hat a({\vec k}_1), {\hat a}^{\dagger}({\vec k}_2)] = \delta^3({\vec k}_1 - {\vec k}_2).
 \label{a1}
 \eea

\noindent with $\omega_k = \sqrt{{\vec k}^2 + m^2\, c^2}$. Instead of the plane waves $e^{\pm i\, (\omega_k\, \tau - \vec k \cdot \vec \sigma)}$ one can use any other basis of positive- and negative-energy solutions of the classical Klein-Gordon equation. By using the creation operators ${\hat a}^{\dagger}(\vec k)$ one can build the standard Fock space starting from the vacuum (defined by $\hat a(\vec k)\, | 0 > = 0$): it describes the particle (or better quanta)
states of the theory.

\medskip

Let us consider a 1-particle state ${\hat a}^{\dagger}(\vec k)\, | 0 >$ with an associated positive-energy solution
$g(\tau, \vec \sigma) = < \vec \sigma | g(\tau) >$ of the classical Klein-Gordon equation. Therefore this wave function
satisfies $i \hbar {{\partial}\over {\partial\, \tau}}\, g(\tau, \vec \sigma) = + \sqrt{m^2\, c^2 - \hbar^2\, {{\partial^2}\over {\partial\, {\vec \sigma}^2}}}\, g(\tau, \vec \sigma)$. Let $< g(\tau) | {\hat {\vec \sigma}} | g(\tau) >$ and $< g(\tau) | {\hat {\vec \pi}} = i \hbar\, {{\partial}\over {\partial\, \vec \sigma}} | g(\tau) >$
denote the expectation values of the position and momentum operators in this state.
\medskip

As shown in Ref.\cite{18} \footnote{See the expanded version 1 of the arXiv paper.} we can consider the multipolar expansion of the wave function $g(\tau, \vec \sigma)$ around a classical trajectory ${\vec \sigma}_{cl}(\tau)$. For all the wave functions with vanishing dipole moment with respect to the classical trajectory we get $< g(\tau) | {\hat {\vec \sigma}} | g(\tau) > = {\vec \sigma}_{cl}(\tau)$ and the Ehrenfest theorem implies ${{d}\over {d\, \tau}}\, < g(\tau) | {\hat {\vec \sigma}} | g(\tau) > = {{d\, {\vec \sigma}_{cl}(\tau)}\over {d\, \tau}} = < g(\tau) | {{{\hat {\vec \pi}}}\over {\sqrt{m^2\, c^2 + {\hat {\vec \pi}}^2}}}| g(\tau) > $ and $ {d\over {d\, \tau}}\, < g(\tau) | {\hat {\vec \pi}} = i \hbar\, {{\partial}\over {\partial\, \vec \sigma}} | g(\tau) > = 0$, so that the classical trajectory is determined by the equation ${{d^2\, {\vec \sigma}_{cl}(\tau)}\over {d \tau^2}} = 0$.
Therefore it is possible to associate an effective particle following an effective mean trajectory only to all the 1-particle states whose wave function has a vanishing dipole.

\bigskip

Already in Minkowski space-time, without going to curved space-times and remaining in the area of condensed matter, the definition of an interacting theory governed by a unitary time evolution is a non-trivial problem: see Ref.\cite{38} for the difficulties to define a self-adjoint Hamiltonian operator bounded from below in free and non-free QFT. Even when this can be done, like in some cases with Hamiltonians bilinear in the creation and annihilation operators, the time evolution implies a unitary (i.e. of the Hilbert-Schmidt type) Bogoliubov transformation leading to new creation and annihilation operators linear combination of the old ones. At each time the instantaneous annihilation operator defines a new instantaneous vacuum, from which a new instantaneous Fock space with a different notion of particle can be created. The new 1-particle states are a superposition of all the states (with every possible particle number) of the initial Fock space. The big open problem is wh
 at kind of quanta (either the initial or the final ones) materialize as effective particles detected by the measuring apparatus (it too can be either inertial or accelerated).

\medskip

In the free case in Minkowski space-time one can consider uniformly accelerated observers (the Rindler ones used for obtaining the Unruh effect \cite{39}): they use a different time-like Killing vector for defining the notion of positive energy and their description of the free Klein-Gordon quantum field is connected with the standard description given by an inertial observer by a Bogoliubov transformation leading to a representation of the free field unitarily inequivalent to the inertial one. Again which one of the unitarily inequivalent quanta give rise to an effective particle to be detected \cite{40} ? The use of Rindler observers for studying the entanglement of modes of the electro-magnetic field in moving cavities in the framework of quantum optics is a quickly developing sector of relativistic quantum information even if the basic interpretational problems are unsolved.

\medskip

Moreover it has been shown \cite{41} that if one describes the free massive Klein-Gordon field in non-inertial frames (like it is done in the Tomonaga-Schwinger formulation on arbitrary space-like hyper-surfaces of Minkowski space-time) then generically the time evolution is not unitarily implementable (the implied Bogoliubov transformation is not of the Hilbert-Schmidt type).

\medskip

In conclusion the notion of particle in QFT is essentially valid for the in- and out-states of the S matrix in inertial frames, a framework relying upon a perturbation
 expansion with suitable ultra-violet and infra-red cutoffs.

\section{Relativistic Entanglement versus Non-Relativistic Entanglement}

After the localization problem let us now look at a simple two-body problem
to display the \textit{changes in its separability and entanglement
properties} going from the non-relativistic case to the relativistic one.
This example will also show the explicit construction of the relativistic
collective variables in the two-body case.\bigskip

As shown in Ref.\cite{42}, the electron-proton system (with masses $m_e$ and
$m_p $ respectively) in the hydrogen atom, governed by the Hamiltonian $H = {%
\frac{{{\vec p}^2_e}}{{2\, m_e}}} + {\frac{{{\vec p}^2_p}}{{2\, m_p}}} - {%
\frac{{e^2}}{{|{\vec x}_e - {\vec x}_p|}}} = H_{com} + H_{rel}$ with $%
H_{com} = {\frac{{{\vec p}^2}}{{2\, M}}}$ and $H_{rel} = {\frac{{{\vec p}^2_r%
}}{{2\, \mu}}} - {\frac{{e^2}}{r}}$ ($M = m_e + m_p$, $\mu = {\frac{{m_e\,
m_p}}{M}})$, can be presented in two ways. Either it is composed by the
subsystems electron $m_e$ and proton $m_p$ (with coordinates and momenta ${%
\vec x}_e$, ${\vec p}_e$ and ${\vec x}_p$, ${\vec p}_p$, respectively) or by
the subsystems center of mass $M$ (with coordinate and momentum $\vec x = {%
\frac{{m_e\, {\vec x}_e + m_p\, {\vec x}_p}}{M}}$, $\vec p = {\vec p}_e + {%
\vec p}_p$) and relative motion $\mu$ ( with coordinate and momentum $\vec r$%
, ${\vec p}_r$). At the quantum level all the positions and momenta become
self-adjoint operators. \medskip

Since both in scattering and bound state theories one describes the dynamics
in the \textit{preferred} momentum basis with given conserved total momentum
$\vec{p}$, the center-of-mass position $\vec{x}$ is un-determined. In the
theoretical description of these theories one never considers wave packets
in $\vec{p}$ with defined localization properties of the center of mass, but
only plane wave with the given value of $\vec{p}$, following the standard
approach without considering the problems of unsharp states  exposed in
the previous Section. \medskip

Therefore a stationary solution of the Schroedinger equation for the
hydrogen atom (it factorizes only in the center-of-mass and relative
variables) in the coordinate representation is

\beq
\psi (\vec{x},\vec{r}) =\phi _{int}(\vec{r})\,e^{{\frac{i}{{\hbar }}}\,%
\vec{p}\cdot \vec{x}}=
\phi _{int}({\vec{x}}_{e}-{\vec{x}}_{p})\,e^{{\frac{i}{{\hbar }}}\,\vec{p}%
\cdot {\frac{{m_{e}\,{\vec{x}}_{e}+m_{p}\,{\vec{x}}_{p}}}{M}}}{\buildrel {def}\over {=}}
\Psi ({\vec{x}}_{e},{\vec{x}}_{p}),
 \label{1}
\eeq

\noindent where $\phi_{int}(\vec r)$ is one of the energy levels of the
atom.  Our presentation here is in the spirit of formal scattering
theory done with plane waves in virtually all text books. In actualilty
the effective particle beams must be described with Gaussian wave packets
with a "classical mean momentum" obtained with flight time methods. \medskip

If we now trace out the center of mass, we get the reduced density matrix $%
\rho_{rel}(\vec r, {\vec r}^{^{\prime }}) = \phi_{int}(\vec r)\,
\phi^*_{int}({\vec r}^{^{\prime }})$ with the associated entanglement
properties of the subsystem \textit{relative motion}. If instead we trace
out the proton, we get the reduced density matrix for the entanglement
properties of the subsystem electron. In Ref.\cite{42} it is shown that it
has the form

\begin{eqnarray}
\rho_{el}({\vec x}_e, {\vec x}_e^{^{\prime }}) &=& \int d^3x_p\, \psi({\vec x%
}_e, {\vec x}_p)\, \psi^*({\vec x}_e^{^{\prime }}, {\vec x}_p) =  \nonumber
\\
&=& e^{ {\frac{i}{{\hbar}}}\, {\frac{{m_e}}{M}}\, \vec p \cdot ({\vec x}_e -
{\vec x}_e^{^{\prime }}) }\, \rho_{int}({\vec x}_e - {\vec x}_e^{^{\prime
}}),
  \label{2}
\end{eqnarray}

\noindent and this implies that it is equally like to observe the electron
in any position since the center of mass is un-determined.

\bigskip

To avoid the complications of the full particle and field configurations
discussed in Ref.\cite{3}, we will consider the simple two-body system
studied in Ref.\cite{4}, which is described in the framework explained in
the Introduction. If $\vec{z}$, $\vec{h}$, are the frozen Jacobi data of the
relativistic center of mass, at the classical level the rest frame
is defined by the embedding of the intrinsic rest 3-spaces of the 3+1 foliation
 into Minkowski space-time (see Refs.\cite{4,6}; $\tau $ and $\sigma
^{r}$ are radar coordinates and the $W$ index on the embedding refers to the
role of the Wigner rest frame)

\begin{eqnarray}
z_W^{\mu}(\tau, \vec \sigma) &=& Y^{\mu}(\tau) + \epsilon^{\mu}_r(\vec h)\,
\sigma^r,\qquad \epsilon^{\mu}_r(\vec h) = \Big( h_r; \delta^i_r + {\frac{{%
h^i\, h_r}}{{1 + \sqrt{1 + {\vec h}^2}}}}\Big),  \nonumber \\
&&{}  \nonumber \\
Y^{\mu}(\tau) &=& \Big(\sqrt{1 + {\vec h}^2}\, (\tau + {\frac{{\vec h \cdot
\vec z}}{{Mc}}}); {\frac{{\vec z}}{{Mc}}} + (\tau + {\frac{{\vec h \cdot
\vec z}}{{Mc}}})\, \vec h + {\frac{{\vec S \times \vec h}}{{Mc\, (1 + \sqrt{%
1 + {\vec h}^2})}}} \Big),  \nonumber \\
{\tilde x}^{\mu}(\tau) &=& Y^{\mu}(\tau ) + \Big(0; {\frac{{- \vec S \times
\vec h}}{{Mc\, (1 + \sqrt{1 + {\vec h}^2})}}}\Big),
 \label{3}
\end{eqnarray}

\noindent where $Y^{\mu}(\tau)$ is the Fokker-Price center of inertia, ${%
\tilde x}^{\mu}(\tau)$ the canonical center of mass, $M$ the invariant mass
and $\vec S$ the rest spin of the two-body system. The external Poincar\'e
group has the generators $P^{\mu} = M c\, h^{\mu} = M\, c\, \Big(\sqrt{1 + {%
\vec h}^2}; \vec h\Big)$, $J^{ij} = z^i\, h^j - z^j\, h^i + \epsilon^{ijk}\,
S^k$, $K^i = J^{oi} = - \sqrt{1 + {\vec h}^2}\, z^i + {\frac{{(\vec S \times
\vec h)^i}}{{1 + \sqrt{1 + {\vec h}^2}}}}$ (the last term in the boost is
responsible for the Wigner covariance of the 3-vectors in the rest Wigner
3-space $\tau = const.$). \medskip

Before adding the rest-frame conditions the world-lines and the 4-momenta of
the two particles are ($V$ is an arbitrary action-at-a-distance potential
\footnote{%
In the electromagnetic case of Ref.\cite{3} the Coulomb potential plus the
Darwin one are outside of the square root.})

\begin{eqnarray}
x^{\mu}_i(\tau) &=& z^{\mu}_W(\tau, {\vec \eta}_i(\tau)) = Y^{\mu}(\tau) +
\epsilon^{\mu}_r(\tau)\, \eta^r_i(\tau),  \nonumber \\
p_i^{\mu}(\tau) &=& h^{\mu}\, \sqrt{m_i^2\, c^2 + {\vec \kappa}_i^2(\tau) +
V(({\vec \eta}_1(\tau) - {\vec \eta}_2(\tau))^2)} - \epsilon_r^{\mu}(\vec
h)\, \kappa_{ir}(\tau),  \nonumber \\
&& |p_i^2| = m_i^2\, c^2 + V(({\vec \eta}_1(\tau) - {\vec \eta}_2(\tau))^2).
 \label{4}
\end{eqnarray}

This equations imply that the un-physical Wigner-covariant 3-positions and
3-momenta inside the rest Wigner 3-space are ${\vec \eta}_i(\tau)$, ${\vec
\kappa}_i(\tau)$, $i = 1,2$.. The conserved internal Poincar\'e generators
are ($M c$ is the Hamiltonian for the motion inside the Wigner 3-space)

\begin{eqnarray}
M\, c &=& \sum_{i=1}^2\, \sqrt{m_i^2\, c^2 + {\vec \kappa}^2_i(\tau) + V(({%
\vec \eta}_1(\tau) - {\vec \eta}_2(\tau))^2)},  \nonumber \\
{\vec {\mathcal{P}}} &=& \sum_{i=1}^2\, {\vec \kappa}_i(\tau) \approx 0,
\nonumber \\
\vec S &=& \sum_{i=1}^2\, {\vec \eta}_i(\tau) \times {\vec \kappa}_i(\tau),
\nonumber \\
{\vec {\mathcal{K}}} &=& - \sum_{i=1}^2\, {\vec \eta}_i(\tau)\, \sqrt{
m_i^2\, c^2 + {\vec \kappa}_i^2(\tau) + V(({\vec \eta}_1(\tau) - {\vec \eta}
_2(\tau))^2)} \approx 0.
  \label{5}
\end{eqnarray}

The rest-frame conditions ${\vec{\mathcal{P}}}\approx 0$, ${\vec{\mathcal{K}}
}\approx 0$, imply that the physical canonical variables in the rest 3-space
are $\vec{\rho}(\tau)={\vec{\eta}}_{1}(\tau) - {\vec{\eta}}_{2}(\tau)$ and $%
\vec{\pi}(\tau) = {\frac{ m_{2}}{M}}\,{\vec{\kappa}}_{1}(\tau) - {\frac{m_{1}%
}{M}}\,{\vec{\kappa}}_{2}(\tau)$, ($M=m_{1}+m_{2}$). Using these relative
variables and imposing the rest frame condition gives the following
expressions for the internal center of mass $\vec{\eta}(\tau)$ (conjugate to
$\mathcal{\vec{P}\approx }0$) and for the world-lines

\begin{eqnarray}
\vec{\eta}(\tau) &=&{\frac{{m_{1}\,{\vec{\eta}}_{1}(\tau) + m_{2}\,{\vec{\eta%
}}_{2}(\tau)}}{M}} \approx {\frac{{m_{1}\,\sqrt{m_{2}^{2}\,c^{2}+H(\tau)}%
-m_{2}\,\sqrt{m_{1}^{2}\,c^{2}+H(\tau)}}}{{M\,(\sqrt{m_{1}^{2}\,c^{2}+H(\tau)%
}+\sqrt{m_{2}^{2}\,c^{2}+H(\tau) })}}}\,\vec{\rho}(\tau),  \nonumber \\
{}&&\qquad H(\tau)={\vec{\pi}}^{2}(\tau)+V({\vec{\rho}}^{2}(\tau)),
\nonumber \\
{} &&  \nonumber \\
&\Downarrow &  \nonumber \\
{} &&  \nonumber \\
M\,c &\approx &\sqrt{m_{1}^{2}\,c^{2}+H(\tau)}+\sqrt{m_{2}^{2}\,c^{2}+H(\tau)%
},\qquad \vec{S}\approx \vec{\rho}(\tau)\times \vec{\pi}(\tau),  \nonumber \\
x_{1}^{\mu }(\tau) &\approx &Y^{\mu }(\tau)+\epsilon _{r}^{\mu }(\vec{h})\,{%
\frac{\sqrt{m_{2}^{2}\,c^{2}+H(\tau)}}{{Mc}}}\,\rho ^{r}(\tau),  \nonumber \\
x_{2}^{\mu }(\tau) &\approx &Y^{\mu }(\tau)-\epsilon _{r}^{\mu }(\vec{h%
})\,{\frac{\sqrt{m_{1}^{2}\,c^{2}+H(\tau)}}{{Mc}}}\,\rho ^{r}(\tau).
 \label{6}
\end{eqnarray}

Let us remark that only in the global inertial frame defined by $\vec{h}=0$
(which we designate as the center-of-mass frame) one has $x_{1}^{o}(\tau
)=x_{2}^{o}(\tau )=\tau $. For any other value of $\vec{h}$ (for instance in
the laboratory frame ${\vec{p}}_{2}(\tau )=0$, so that $\vec{\pi}$ becomes
parallel to $\vec{h}$) the time variables of the world-lines do not
coincide, so that we cannot make equal time statements by using them (for
more details see the relativistic kinetic theory of fluids developed in Ref.%
\cite{43}). \medskip

The quantization of the model is done in the \textit{preferred} $\vec h$%
-base. The variables $\vec h$, $\vec \rho$, $\vec \pi$, are replaced by
self-adjoint operators. The open problems are whether we replace $\vec z$
with either a self-adjoint or a symmetric operator (see the discussion on
the Newton-Wigner operator in Section III) and whether we accept either only
momentum plane waves or also wave packets with some localization of the
center of mass (the particle beams with mean classical trajectory). If $\vec
z$ becomes a self-adjoint operator, then also the external Lorentz
generators can be made self-adjoint after a suitable ordering. As a
consequence, the operators corresponding to the world-lines of the particles
become complicated objects needing non-trivial orderings except in the case $%
\vec h = 0$, the only one in which $\vec \rho = {\vec \eta}_1 - {\vec \eta}%
_2 = {\vec x}_1 - {\vec x}_2$.

\medskip

If we work in the $\vec{h}$, $\vec{\rho}$, basis of the Hilbert space and we
fix $\vec{h}=\vec{k}$, then the wave function is $\psi _{\vec{k}}(\vec{h},%
\vec{\rho},\tau )=\delta ^{3}(\vec{h}-\vec{k})\,\phi (\vec{\rho},\tau )$
with $\phi (\vec{\rho},\tau )$ satisfying the Schroedinger equation $%
i\,\hbar \,{\frac{{\partial }}{{\partial \,\tau }}}\,\phi (\vec{\rho},\tau )=%
{\hat{M}}\,c\,\phi (\vec{\rho},\tau )$. By putting $\phi (\vec{\rho},\tau
)=\exp (-{\frac{i}{\hbar }}\,{\frac{\epsilon }{c}}\,\tau )\phi (\vec{\rho})$%
, the stationary solutions $\phi _{nlm}(\vec{\rho})$ satisfy the equations

\begin{eqnarray}
&&\hat M\, c\, \phi_{nlm}(\vec \rho) = \epsilon_n\, \phi_{nlm}(\vec \rho),
\nonumber \\
&&{\hat {\vec S}}^2\, \phi_{nlm}(\vec \rho) = l\, (l + 1)\, \phi_{nlm}(\vec
\rho),\qquad {\hat S}_3\, \phi_{nlm}(\vec \rho) = m\, \phi_{nlm}(\vec \rho),
\label{7}
\end{eqnarray}

\noindent and the external 4-momentum becomes $P^{\mu}_n = {\frac{1}{c}}\, %
\Big(\epsilon_n\, \sqrt{1 + {\vec k}^2}; \epsilon_n\, \vec k\Big)$. In the $%
\vec z$ basis of the Hilbert space we get a plane wave $e^{{\frac{i}{\hbar}}%
\, \vec k \cdot \vec z}$ for the delocalized center of mass. \medskip

Regarding entanglement we can trace out the center of mass and find the
reduced density matrix of the subsystem "relative motion": it is of the type
$\rho_{rel}(\vec \rho, {\vec \rho}^{^{\prime }}) = \phi(\vec \rho)\, \phi^*({%
\vec \rho}^{^{\prime }})$ like in the non-relativistic case.

\medskip

In the center-of-mass frame $\vec{h}=0$, where $\vec{\rho}={\vec{\eta}}_{1}-{%
\vec{\eta}}_{2}={\vec{x}}_{1}-{\vec{x}}_{2}$, for $\vec\rho = {\vec \rho}
{}^{\prime }$ we get $\rho _{rel}(\vec{\rho},\vec{\rho})=|\phi (\vec{\rho}%
)|^{2}=\rho _{int}({\vec{x}}_{1}-{\vec{x}}_{2})$ to be compared with the
non-relativistic equation (4.2) with $\vec{p}=0$. \medskip

However we cannot study the subsystem 1 tracing out the subsystem 2: this is
the spatial non-separability of relativistic entanglement discussed in the
Introduction and in Ref.\cite{6}. \medskip

The same problems would appear in the study of the entanglement in
scattering processes: see Refs.\cite{44} for some of the existing results.

\section{Reconciliation of the Relativistic and Non-Relativistic Worlds
Taking into Account the Preparation and the Detection of Particles in
Experiments}

We have seen that the non-relativistic notion of separability, according to
which the Hilbert space of a quantum system composed of subsystems is the
tensor product of the Hilbert spaces of the subsystems (zeroth postulate of
NRQM), is destroyed in special relativity where clock synchronization is
needed to define 3-space and to avoid relative times in bound states.. This
fact, together with the non-local nature of relativistic collective
variables, induces a spatial non-separability implying that we can speak of
subsystems only at an un-physical level, that existing before adding the
rest-frame conditions. After their imposition we describe the overall system
by a non-measurable external canonical non-covariant decoupled center of
mass and by an internal world of Wigner-covariant relative variables. Only
the frozen Jacobi data of the center of mass condition $\vec{h}=0$ and the
relative variables can be quantized consistently at the physical level.

\medskip

This picture is strongly different from the standard non-relativistic
framework, where there is unitary equivalence between the presentation with
separable subsystems and the one with Newton center of mass and relative
variables. However we have seen in Section III that there are problems also
with the localization of the Newton center of mass (unsharp positions).

\medskip

Let us consider a non-relativistic (but the same is true at the relativistic
level) experiment testing a quantum system, let us say a two-particle
system. There are dynamical observers (replaced with mathematical ones like
Alice and Bob) using some apparatus for the preparation of the beam of
particles to be used to define the system, having the particles interacting
in a well (classically defined) way and using some detectors to extract
information from the process. As said in the Introduction all these steps,
realized by the observers, are imagined and realized by using a strongly
classical intuition, in agreement with Bohr's point of view according to
which every feasible experiment must admit a classical description.

\medskip

Coming back to the experiment, we see that the two incoming particles of our
system are in special states (mean trajectory and mean momentum) prepared by
the apparatus and that the outgoing particles can be detected only if they
have a mean trajectory collineated with the detectors. \medskip

This means that at the experimental level an isolated two-particle system is
a mathematical idealization. As said in the Introduction at best it is a
(non-unitary evolving) \textit{open quantum system} \cite{17} always with
some interaction first with the preparing apparatus and then with the
detectors, both of which should be considered as quantum many-body systems.
\medskip

Moreover, as noted in Ref.\cite{45}, the observers are not able to define a
perfect classical mathematical reference frame, but only a \textit{bounded}
one determined by the level of precision of every instrument used. \medskip

As a consequence, in a realistic description of an experiment in NRQM one
should consider as an idealized isolated quantum system at least the union
of our two-particle system plus an environment composed by a quantum
many-body description of the preparing apparatus and of the detecting one
(they admit an effective classical description in terms of emerging
effective notions like a pointer). The presence of interactions forces us to
work in the position basis of the overall Hilbert space by using the
decoupled Newton center of mass of the whole system and relative variables.
Even if the non-relativistic center-of-mass position operator would be
measurable (and this is an open problem due to unsharp positions), it is out
of reach of the experiment (it pertains to the reference frame of the
observer). What is measured of the particles are their relative variables
with respect to the preparing apparatus and, after the interaction in the
experimental area, with respect to the detectors (the effective mean
trajectories of the incoming and outgoing particles) plus the relative
variable between the two particles (it controls their mutual interaction).
This is in accord with the experimental results of Ref.\cite{33}: only the
relative positions of atoms are measurable.

For all practical purposes (FAPP) this description is the same that we have
at the relativistic level, now both for the isolated two-body problem and
for the system particles plus experimental apparatus.

\medskip

Therefore at the experimental level there is not a drastic difference
between the relativistic and the non-relativistic frameworks induced by the
Lorentz signature of Minkowski space-time below the threshold of pair production.

\section{Concluding Remarks}

We have emphasized the differences between relativistic and non-relativistic
quantum mechanics and the associated notions of entanglement in inertial
frames and at the same time revealed unexpected common features.

\medskip

Due to the Lorentz signature of Minkowski space-time, creating the problems
of clock synchronization to define 3-space, of the elimination of the
relative times in bound states and of the non-uniqueness of the relativistic
collective variables, at the relativistic level we have global spatial
non-separability limiting the existence of subsystems to an un-physical
level before adding the rest-frame conditions to eliminate the internal
collective variable in the Wigner 3-space. The external decoupled canonical
non-covariant relativistic center of mass is a non-local, and therefore
non-measurable, quantity already at the classical level. \medskip

These non-separability and non-locality both at the classical and quantum
levels reduce the relevance of the still debated quantum non-locality of
NRQM. As shown in Ref.\cite{3} these properties are present also in CFT and their extension to QFT
is a difficult open problem, which adds to the existing problems with the notion of particle in QFT \cite{36} (instability under either unitary or non-unitary Bogoliubov transformations).
The relativistic non-separability and non-locality point towards non-local QFT
and rise the problem of the validity of \textit{micro-causality} (quantum operators at space-like
distances commute) at the relativistic level. This issue has already been
raised by Busch \cite{46} with the notion of \textit{\ unsharp observables}
(a local operator not measurable with local actions in a given 3-region).
According to him sharp spatial localization is an \textit{operationally
meaningless idealization} (it requires an infinite amount of energy with
unavoidable pair production; the quantum nature of the constituents of the
detectors should be taken into account; and so on). \medskip

We have shown the status of the particle localization problem both at the
relativistic and non-relativistic levels showing the existing theoretical
problems and some of the experimental limitations. One would need the
identification of some relevant (well mathematically defined) position basis
of wave functions with compact support (or like the over-complete coherent
state basis for the harmonic oscillator) for the position operator to be
used for every type of localization problem. \medskip

By taking into account the quantum nature of both the preparing apparatus
and of the detectors we have shown that also at the non-relativistic level
the real quasi-isolated system with a decoupled center of mass is the full
set of "apparatus + detectors + observed system" and that the prepared and
detected particles, moving along mean classical Newton trajectories, are
described by relative variables. \medskip

Since, in any case, electromagnetism is always present, this implies that
the relativistic picture is valid at every level in experiments. \medskip

We have shown the \textit{non-factual} nature of the mathematical time-like
observers and of their mathematical synchronization conventions for building
reference frames (see also Ref.\cite{47} in the framework of quantum
information). It seems quite difficult to develop a theory in which Alice
and Bob are dynamical observers \footnote{First steps in this direction are done with {\it quantum metrology}
(see Ref.\cite{48}, ch.7), in which a quantum system is described by means of the relative variables with respect to another quantum system (the observer) without using variables defined with respect to an external classical reference frame (this relational approach is consistent with relativistic non-separability). It is under investigation what is the change in the information and in the entanglement if one goes from the description with respect to a quantum observer to the one with respect to another quantum observer.} and exchange information by using a
dynamical electromagnetic field! Instead Alice and Bob are present in every
protocol for relativistic quantum information (see Ref.\cite{49} for an old
review emphasizing the problems with special relativity), a quickly
developing theory, in which the study of the spatio-temporal trajectories of
the investigated quantum systems is nearly always lacking. \medskip

We have nothing to say about the probabilistic Born rule (randomness of the
unique outcomes) and on the nature of quantum reality (objective,
subjective, mixture of classical reality and information theory,...).
However, if the wave function of a quantum system describes its properties
(no ensemble interpretation), then the experimentally observable wave
functions have an associated emergent classical description in terms of
Newton trajectories. Moreover, the spatial non-separability, introduced by
SR already at the classical level, gives rise to many non-ignorable
problems: A) the world-lines of the observers and of the macroscopic
apparatuses are inter-wined with the investigated relativistic (classical or
quantum) system and must be taken into account; B) as already said this
implies that the observers can no longer be treated as mathematical
decoupled entities but must be macroscopic bodies localized in the
space-time; C) the non-separability together with Busch unsharpness show
that causality problems can no longer be solved by saying that systems in
disjoined regions with space-like distance are un-related; D) therefore
foundational statements like the freedom to choose the measurements settings
independently from the investigated quantum system (see for instance Refs.%
\cite{50}) are no longer meaningful.

Finally in this paper we have only considered inertial frames. The extension
of our results to non-inertial frames is now under investigation, with an
attempt to avoid uniformly accelerated Rindler observers (they disappear
with the light-cone in the non-relativistic limit) but taking into account
Unruh-DeWitt detectors \footnote{
See Ref.\cite{51} for the relativistic description of the two-level atoms
used in Unruh-DeWitt detectors.} (see the review paper \cite{22} and the bibliography of Ref.\cite{52}). The
extension of these ideas to classical general relativity to include gravity
can be done along the lines described in the review paper \cite{5}, but the
totally open problem is to find a consistent theory of quantum gravity \cite{37}.

\vfill\eject

\end{document}